\documentclass[12pt]{article}
\usepackage{latexsym}
\usepackage{graphicx}
\newcommand{\beq}{\begin{equation}}
\newcommand{\eeq}[1]{\label{#1}\end{equation}}
\newcommand{\bea}{\begin{eqnarray}}
\newcommand{\eea}[1]{\label{#1}\end{eqnarray}}

\begin{document}
\begin{flushright}
\hfill{IMPERIAL-TP-2014-MJD-06}\\
\end{flushright}
\vspace*{0.37truein}
\centerline{\bf M-history without the M\footnote{Commissioned by Contemporary Physics. DOI:10.1080/00107514.2014.992964}}
 \vspace*{0.37truein} \centerline{\footnotesize M.
J.  Duff\footnote{m.duff@imperial.ac.uk} }\vspace*{0.015truein}
\centerline{\footnotesize\it Blackett Laboratory,}
\baselineskip=15pt
\centerline{\footnotesize\it  Imperial College London, SW7 2AZ, UK}
\bigskip


\vspace{20pt}

\abstract{ A review of the book ``A Brief History of String Theory: From Dual Models to M-Theory'' by Dean Rickles.
\bigskip\bigskip
\indent

Prior to the 1984 superstring revolution,  most theoretical physicists paid little attention to the work of Green and Schwarz on ten-dimensional superstrings. Many preferred  supergravity in eleven dimensions, the maximum dimension permitted by supersymmetry of the elementary particles.  So one can sympathise with John Schwarz when he complains in the book that this pioneering work went largely unrecognised until 1984.

However, most string theorists were in their turn slow to recognise the importance of membranes and eleven dimensions.  There are no superstrings in eleven dimensions but, as was shown in 1987, there are supermembranes \cite{Bergshoeff:1987cm,Duff:1990xz,Duff:1999}, which is why between 1984 and 1995 many string theorists were opposed to eleven dimensions.  Membrane-related grant proposals tended to attract hostile referee reports during that period and papers with titles like ``Supermembranes: a fond farewell''  did not help.  One string theorist announced that ``I want to cover up my ears every time I hear the word membrane'' and some organisers of the annual superstring conferences even banned the use of the ``M-word''.  My colleague Paul Townsend, one of the membrane pioneers, compared this with the theatrical superstition of calling MacbethÕ the ``M-Play''. This opposition continued even after it was shown in 1987 that one of the five consistent ten-dimensional superstring theories, the Type IIA string, was just the limiting case of the eleven-dimensional supermembrane \cite{Duff:1987bx} and even after it was shown in 1994 that the spectrum of states that resulted from compactifying the  membrane theory from eleven dimensions to four was identical to that resulting from compactifying  the Type IIA string from ten dimensions to four \cite{Hull:1994ys,Townsend:1995kk}.

However, attitudes to eleven dimensions among both string and membrane theorists underwent a sea-change in March 1995 when string-guru Edward Witten, drawing on work by Hull and Townsend, Sen and myself, made an astonishing announcement at the Strings 95 conference in Los Angeles. The five consistent strings (1, IIA, IIB, HE, HO) were not, as previously thought, five rival candidates for the final theory but were merely five corners of a deeper and more profound eleven-dimensional structure \cite{WikiM}, which he called M-theory. The ultraviolet divergences of $D=11$ supergravity are now seen to be irrelevant because it is just the low-energy approximation to the underlying M-theory. Curiously, however, Witten did not immediately embrace membranes even then: 

\medskip
\noindent
{\it There is, for instance, no evidence for membrane excitations; such evidence might well have appeared if a consistent membrane theory with eleven-dimensional supergravity as its low energy limit really does 
exist \cite{Witten:1995ex}.}

\medskip
\noindent
{\it It has been proposed that the eleven-dimensional theory is a supermembrane theory but there are some reasons to doubt that interpretation, we will non-committally call it M-theory, leaving for future the relation of M to membranes \cite{Horava:1995qa}.}

\medskip
\noindent
However, it soon became clear  that the equations of M-theory in diverse dimensions less than or equal to eleven admit solutions describing a variety of ``$p$-branes'',  with a web of ``dualities'' relating them. (Here a $0$-brane is a point-particle, a $1$-brane is a string, a $2$-brane is  a membrane and so on.)  Prior to these developments, string calculations had to rely on ``perturbation theory'', an approximation that involves keeping just the first few terms in an expansion in powers of some small parameter $g$. But the importance of these dualities was that they permitted for the first time non-perturbative calculations, relating a theory with parameter $g$ to another with parameter $1/g$. While remaining open-minded about the fundamental degrees of freedom of M-theory,  Witten was soon converted to the utility of membranes and went on to make contributions to membrane theory as profound as those he had made to string theory.

October 1995 marked another major development in the history of membranes and M-theory when Joe Polchinksi \cite{Polchinski:1995mt} observed that a subset of the $p$-brane solutions to  $D=10$ Type IIA and Type IIB supergravities admit the alternative interpretation of Dirichlet-branes (or $Dp$-branes): surfaces of dimension $p$ on which open strings can end. Importantly for the purposes of this review, since D-branes may be framed within the language  of ten dimensional superstrings they were psychologically more palatable to those who had opposed eleven dimensional membranes which were then renamed M-branes. Thus the ``electric'' $D=11$ membrane becomes the M2-brane, and its ``magnetic'' dual the M5-brane.

D-branes had profound implications.
For example, when combined with the M-branes, they led to Juan Maldacena's $AdS_d/CFT$ correspondence which conjectured a remarkable equivalence between gravitational theories living in $d$-dimensional spacetimes of the saddle-shaped anti-de Sitter (AdS) variety and non-gravitational conformal field theories (CFT) living on their ($p=d-2$)-dimensional boundaries. Maldacena's paper, which focussed on $AdS_4/M2$-branes, $AdS_5/D3$-branes and $AdS_7/M5$-branes, has garnered an incredible $>10,000$ citations. D-branes also figure prominently in the M-theoretic microscopic derivation of Bekenstein-Hawking black hole entropy and also in the ``brane-world'' which supposes our universe to be a 3-brane floating in a higher-dimensional bulk space-time

So while acknowledging their oversight in ignoring strings before 1984, those working  on eleven dimensional supergravity from 1979  and supermembranes from 1986 need offer no apology for doing so since branes and eleven dimensions are both vital ingredients of M-theory. They should not be too bothered if their work on membranes went largely unrecognised until 1995; if the membrane/string theorists were unable to convince their strings-only colleagues, perhaps they were partly to blame. Nor should they mind that at international conferences those same colleagues would slap them on the back and say ``So, what's it like to be working in the wrong dimension?''

 What might bother them, however, was the attempt by a vocal minority to belittle their work after its importance to M-theory became clear, creating a  smokescreen designed to blur the the 10/11 string/membrane distinctions. This included the pretence that strings live in 11 dimensions after all:

\medskip
\noindent
{\it The way the string vibrates determine each particle's properties. This all takes place in convoluted landscape of 11-dimensional space \cite{Cole1},}

 \medskip
\noindent
  some odd attributions:

\medskip
\noindent
{\it In 1995, Witten sparked the latest revolution, introducing vibrating sheets called membranes and bringing the total number of new dimensions to 11 \cite{Cole2},}

\medskip
\noindent
some grudging concessions:

\medskip
\noindent
{\it But this eleven dimensional theory would not die. It eventually came back to life in the strong coupling limit of superstring theory in ten dimensions \cite{Official}},

\medskip
\noindent
and some (in my opinion) misleading ones:

\medskip
\noindent
{\it What makes a p-brane? A p-brane is a spacetime object that is a solution to the Einstein equation in the low energy limit of superstring theory, with the energy density of the nongravitational fields confined to some p-dimensional subspace of the nine space dimensions in the theory \cite{Official}.}

\medskip
\noindent
Sadly, in ``A Brief History of String Theory: From Dual Models to M-Theory'' Dean Rickles makes it clear he also belongs to this minority. For example:
 
 \medskip
\noindent
 (1) Rickles promises to explore how M-theory came into being, but although D-branes are discussed on pages 4, 7, 208, 212-214, 216, 221, 223, 224, 226, M-branes are nowhere mentioned, hence the title of this review\footnote{ Other similar M-brane-free zones may be found  in Wikipedia's ``History of string theory'' \cite{Wikistring},
in Lisa Randall's book \cite{Randall} ``Warped Passages'', in Lawrence Krauss's book \cite{Krauss} ``Hiding in the Mirror'' and in Sean Caroll's book \cite{Carroll}  ``The Particle at the End of the Universe'' (Is this the Maldacena dual of The Membrane at the End of the Universe \cite{Duff:1988st}?).  Brian Greene, bless his heart, gets it right \cite{Greene1, Greene2}. }.  In particular, the original eleven-dimensional supermembrane papers  \cite{Bergshoeff:1987cm,Duff:1990xz}  are not discussed at all. In his analysis of the Maldacena duality conjecture, for example, Rickles just cherry-picks the D3 and omits all mention of the M2 and the M5 (pages 7, 223-226). All this involves creating the impression on page 7 that Polchinski's 1995 D-brane paper predated Witten's 1995 M-theory paper. 
 
 \medskip
\noindent
 {\it Phase 4 (Beyond Strings) [1995-present] string theory is understood to contain objects, Dp-branes, of a variety of dimensionalities (of which strings are a single example, for p=1). More dualities are introduced, leading to a conjecture that the different string theories (and a further 11-dimensional theory) are simply limits of a deeper theory: M-theory.}
 
  \medskip
\noindent
Thus Rickles echoes the Wikipedia version of the History of String Theory, according to which research on branes began only in 1995. 

 \medskip
\noindent
{\it  In the mid 1990s, Joseph Polchinski discovered that the theory requires the inclusion of higher-dimensional objects, called D-branes \cite{Wikistring}. }

  \medskip
\noindent
 To his credit, Joe Polchinski himself does not make this claim \cite{Rickles}. In any event, telling the full story would not downgrade Polchinski's role in the scheme of things since he co-authored the first paper on supermembranes (in six rather than eleven dimensions) in 1986 \cite{Hughes:1986fa} and was talking about D-branes already at the 1989 Texas A\&M Strings Conference \cite{AM}.

\medskip
\noindent
(2) Rickles calls the membrane/D=11 revolution of 1995 the ``second superstring'' revolution (though he is not alone in this). Accordingly, in the text membranes are called ``higher-order strings''. See page 17. The M-word itself appears only in derisory footnotes on pages 188, 209, 218.
On the origin of the term ``M-theory'', Rickles says on page 217:

\medskip
\noindent
{\it In a later popular article of Witten's we find the oft-quoted explanation of the letter `M': ``M stands for magic, mystery or matrix, according to taste'' \cite{Matrix}.}

\medskip
\noindent
conveniently forgetting Witten's earlier ofter-quoted explanation:

\medskip
\noindent
{\it For instance, the eleven-dimensional ÔM-theoryÕ (where M stands for magic, mystery or membrane, according to taste) \cite{Witten:1995em}.}
\medskip
\noindent

\medskip
\noindent

\medskip
\noindent
(3) Rickles finds it necessary to belittle the role of supergravity compared with superstrings in the historical development of M-theory, calling the years between the discovery of supergravity and the superstring revolution the ``Decade of Darkness''. While it is true that eleven-dimensional quantum supergravity suffers from the ultraviolet divergences that ten-dimensional superstrings avoid, its very existence calls into question the notion that strings are the be-all-and-end-all of the final theory.  Thus my 1987 paper ``Supermembranes: The First Fifteen Weeks''\footnote{Which Rickles mis-cites twice as `` fifteen days'' and ``fifteen years''} begins:
  
\medskip
\noindent
{\it Many of the supergravity theories that we used to study a few years ago are now known to be merely the field theory limit of an underlying string theory.  What are we to make, therefore, of supergravity theories which cannot be obtained from strings such as N = 1 supergravity in eleven dimensions?\cite{Duff:1987qa}}

\medskip
\noindent
Yet in his zeal to downgrade supergravity Rickles distorts the compliment to make it sound more like an insult:

\medskip
\noindent
{\it This became widely accepted, and one can find Michael Duff writing in 1988 that ``Many of the supergravity theories that we used to study a few years ago are now known to be merely the field theory limit of an underlying string theory''}
\medskip
\noindent

It is too early to tell whether superstring theory will be vindicated experimentally as a description of elementary particles. In my opinion, however, its reconciliation of gravity and quantum mechanics is one of the intellectual triumphs of the  twentieth century. As such, its historical standing is not diminished by acknowledging that M-theory also owes a debt to supergravity and supermembranes. As a historian, Rickles is no doubt familiar with the phrase ``premature anti-nazis'' aimed at those Russians who opposed the pre-war Molotov-Ribbentrop pact after became clear they were right all along to do so. In 
Rickles' s view of history, those who advocated branes and eleven dimensions in between the 1984 string revolution and the 1995 M-revolution occupy a similar status. String theory  deserves better, in my opinion. 

\bigskip
\bigskip
ADDED NOTE: After this review was accepted for publication, Polchinski posted a paper \cite{Polchinski:2014mva} written for a special issue of Studies in History and Philosophy of Modern Physics.  This provides a more up-to-date account of Polchinski's version of brane history than \cite{Rickles}.  A similar version was subsequently posted by Hubeny \cite{Hubeny:2014bla} for a special issue of Classical and Quantum Gravity  on ``Milestones of General Relativity".

On the subject of D-branes in the pre-1995 era, to the best of my knowledge, the D3-brane (and the M5-brane) first appeared in the 1987 conformal ``brane-scan'' of Table I in \cite{Blencowe:1987} obtained by regarding super p-branes as occupying the boundaries of AdS spacetimes \cite{Duff:1987qa,Duff:1988st}.  It was pointed out in 1991 that the D3 worldvolume Lagrangian is that of a four-dimensional N=4 supersymmetric gauge 
theory \cite{Duff:1991pea}.

\end{document}